\documentclass{article}%book(ctexbook), report(ctexrep), letter
\usepackage{amsmath}
\usepackage{extarrows}
\usepackage{supertabular}
\usepackage[x11names]{xcolor}
\usepackage{graphicx}
\graphicspath{{graphs/}} % 指定子文件夹路径
\usepackage{tikz}
\usepackage{amssymb}
\usepackage{mathrsfs}

\usepackage{float}
\usepackage{caption}
\usepackage{subcaption} % 在导言区添加
\usepackage{nicematrix}
\usepackage{multirow}

\usepackage[backref]{hyperref}
\usepackage{geometry}

\usepackage{appendix}
\usepackage{authblk}

\newcommand{\md}{\mathrm{d}}

\title{Quasinormal modes of an electrically charged Kalb-Ramond black hole}
\author[1,2]{Yun-Tao Gu}
\author[1,2]{Wen-Di Guo}
\author[1,2]{Yu-Xiao Liu\thanks{Corresponding author. Email: liuyx@lzu.edu.cn}} % 用thanks标记通讯作者

\affil[1]{
	Lanzhou Center for Theoretical Physics,\protect\\
	Key Laboratory of Theoretical Physics of Gansu Province,\protect\\
	Key Laboratory for Quantum Theory and Applications of the Ministry of Education,\protect\\
	Gansu Provincial Research Center for Basic Disciplines of Quantum Physics,\protect\\
	Lanzhou University, Lanzhou 730000, China
}
\affil[2]{
	Institute of Theoretical Physics \& Research Center of Gravitation,\protect\\
	School of Physical Science and Technology,\protect\\
	Lanzhou University, Lanzhou 730000, China
}
\date{}

\begin{document}
	\maketitle
	\begin{abstract}
		%Lorentz violation is an important feature of modified theories. Kalb-Ramond field induced spontaneous Lorentz violation has drawn broad attention. Recently, an electrically charged Kalb-Ramond black hole was proposed. A minor error in the expression of energy-momentum tensor is identified. We study the quasinormal modes of the ``undecouplable'' system using the matrix-valued continued fraction method and the matrix-valued direct integration method. A method to recognize the different modes of an ``undecouplable'' system is developed. Error analysis is done and the influence of Lorentz violation on fundamental modes are analyzed within appropriate range of parameter.
		Lorentz violation serves as a significant feature in many modified theories of gravity. In particular, spontaneous Lorentz violation induced by the Kalb-Ramond field has attracted considerable attention. Recently, an electrically charged black hole solution within the Kalb-Ramond framework was proposed. In this study, we investigate the quasinormal modes of the resulting ``undecouplable'' system using both the matrix-valued continued fraction method and the matrix-valued direct integration method. Additionally, we develop a new approach to distinguish between different modes in such ``undecouplable'' systems. An error analysis is performed, and the influence of Lorentz violation on the fundamental quasinormal modes is systematically analyzed within a suitable parameter range.
	\end{abstract}
	
\section{Introduction}
	Finding the unified theory, which is usually suggested to be a quantum gravity theory, is an essential issue in theoretical physics. Various possibilities have been proposed, among which the most promising one seems to be string theory. Modified gravity theories serve as low-energy effective theories of the underlying quantum gravity. One candidate for Planck-scale signals emerging from the underlying unified quantum gravity theory is Lorentz violation (LV).
	
	%Quantum gravity are expected to occur at Plank scale, so low-energy tests are needed. Not to mention with so many alternatives, these tests are necessary for a phenomenological construction from bottom to top as well. Even when considering different high-energy theories, similar low-energy effects, e.g., holography, are expected to occur as general features of quantum gravity theories.
	
	%Modified gravity theories serve as low-energy effective theories of the underlying quantum gravity, in which the features mentioned above are described by phenomenological terms regardless of their origins.
	
	Among theories incorporating LV, explicit LV is a significant category. Its defining characteristic is the presence of vacuum expectation values (VEVs) that remain invariant under active (or particle) Lorentz transformations. A geometrical framework allowing for nonzero vacuum quantities, i.e., the VEVs, that violate local Lorentz invariance but preserve general coordinate invariance is therefore required. Riemann-Cartan geometry is particularly well-suited for this purpose~\cite{Cartan-LV}. The Standard-Model Extension (SME) serves as a phenomenological effective field theory that describes explicit LV~\cite{explicit-LV}. 
	Notably, an interesting property to mention is that LV in flat spacetime is closely connected with CPT violation~\cite{CPT-LV1,CPT-LV2}. In the Minkowski-spacetime limit of the SME, the Lorentz-violating terms can be classified according to their properties under CPT.
	
	One attractive and generic mechanism of explicit LV is spontaneous LV~\cite{spontaneous-LV}. It was first proposed in string theory and was considered to be a distinctive feature of string theory---unlikely to occur in four-dimensional renormalizable gauge theories. The basic idea is that, a field with negative mass-squared term possesses an unstable vacuum. In string field theory, tensor fields obtain such negative mass-squared terms through cubic interaction of the form $ST^MT_M$ with the scalar field $S$ having a negative VEV, such as the tachyon field. Here, the index $M$ denotes one or more Lorentz indices.
	Another mechanism giving rise to explicit LV is noncommutative field theory, which is equivalent to a subset of the SME~\cite{NFT}. The commutator of spacetime coordinates is said to have the properties as the VEVs of fields in the string spectrum. Such a structure occurs naturally in string theory~\cite{NFT-geometry}. Furthermore, Ref.~\cite{NFT-bi} demonstrates that birefringence effects are allowed in this theory.
	%Noncommutative field theory~\cite{NFT} comes from the idea that the spacetime coordinates being intrinsically noncommutative. The commutator tensor is said to have the properties as the VEVs of fields in the string spectrum. It is found to be equivalent to a subset of the SME. And since the revival of this theory was because it's natural occurrence in sting theory~\cite{NFT-geometry}, it gives us more confidence to say that it's maybe proper to classify this kind of theoretical origin with the spontaneous Lorentz violation discussed above in string theory.
	
	It has also been suggested that certain theories such as loop quantum gravity~\cite{Loop-LV} and multiuniverses~\cite{emergent} might generate birefringence effects~\cite{Loop-LV,braneworld} that provide evidence for LV. However, the nature of LV in these frameworks differs intrinsically from the Lorentz gauge symmetry violation. Moreover, it is not caused by VEV-type coefficients for LV. Instead, LV in loop quantum gravity reflects the feature of the chosen semi-classical state rather than an inherent feature of the fundamental theory~\cite{Loop-LV,weave}, while LV in multiuniverses is more likely to be assuming a complete breakdown of Lorentz invariance~\cite{emergent,multiverse,emergent-condensed}. Given that this paper focuses on the spontaneous violation case, our analysis should not be regarded as representative of all possible forms of LV.

	A vector-induced model, now commonly referred to as the Bumblebee model, was first proposed in Ref.~\cite{gravitational-LV} and has drawn wide interest. The name “Bumblebee” was subsequently introduced by the same author in Ref.~\cite{explicit-LV}. A frequently used picture for understanding LV in this particular model emphasizes that the VEV of the vector field implies a preferred direction at each point in spacetime. It is also worth noting that the theory imposes constraints on compactification of the extra dimensions~\cite{gravitational-LV}. However, the Bumblebee model is a minimal framework containing the feature of spontaneous LV. The Bumblebee field might just correspond to any vector field in the first excited state of open bosonic string spectrum. Meanwhile, in this work we turn our attention to a tensor-induced model, in which the term “tensor” refers specifically to the Kalb-Ramond (KR) field.

	The KR field, also known as B-field, was first proposed in the study of interstring interactions~\cite{classical-KR}. In modern textbooks it is described as the antisymmetric 2-tensor that appears alongside graviton and dilaton in the massless excited state in the oriented close string spectrum (see e.g.~\cite{polchinski2005string}). In the context of superstring and heterotic string theories, it arises specifically as part of the NS-NS sector. The ten-dimensional Einstein-Kalb-Ramond action was first presented in Ref.~\cite{action-KR}. In four dimensions, the KR field strength 3-form can be written with its dual 1-form $\mathbf{T}$. One can show that $\mathbf{T} = \md\chi$ is also exact if the first homology group of the spacetime manifold is trivial. The $\chi$ field has long been considered as an axion field despite its tendency toward dynamical instability. And it was suggested by Ref.~\cite{induced-KR} that a Routh's method can solve the problem caused by the noncommutativity between varying the action with respect to the metric field and substituting KR field with the $\chi$ field. Later it was shown that a candidate for the hypothetical axion field is the dual incarnation of KR field in KK-compactifications of heterotic string theory to 4d~\cite{axion-KR-witten-old,axion-KR-witten}.
	
	%最近的一篇文章提出“在广义相对论的拓展中，准正则模不再是等谱的，引力波传播也不再沿测地线且会发生双折射。存在唯一的拉氏量，对引力波有非双折射的色散关系以及在光程函数极限下有等谱的准正则模。作者注意到最低阶广义相对论的等谱修正与来自type II弦论的四次曲率修正一致，表明等谱性可能是量子引力的关键性质。”我查了一下bumblebee-induced和KR-induced的洛伦兹破缺理论中，前者不存在双折射而后者存在双折射，如果那篇文章中的结论正确，即量子引力的关键性质是不存在双折射，那么是否说明大黄蜂引力比KR场引力更靠谱？但是我又注意到KR场与轴子场在某种杂化弦论中存在对偶关系，这是介绍KR场常常提到的，但前面的文章中提到的性质来自type II弦论，这似乎不是杂化弦？那么如果我们认为研究KR场的动机与轴子场有关，是不是就不用考虑前面文章中提到的不存在双折射的限制？
	
	We note that a recent publication argued that nonbirefringence might serve as a key feature of quantum gravity, because the extended action allowing nonbirefringence meets with the effective action of type II theories~\cite{birefringence-string}. Previous research indicates that Bumblebee theory predicts birefringence while parity violation with KR field forbids this effect~\cite{birefringence-Zhu}. However, reading the Refs.~\cite{stringEFT-witten,stringEFT-Gross,stringEFT-new} in detail, although, as discussed in Ref.~\cite{birefringence-string}, the graviton part of the effective Lagrangian given by the four-point scattering amplitude in type II string theory does possess the same form as the extended action analyzed, the latter is not the complete modification unless a field redefinition is performed to incorporate the other fields including the KR field and the dilaton field. Furthermore, even though the effective Lagrangians given by type II theory and heterotic string in this representation do not contain the coupling in our LV Lagrangian, these terms might still appear through a redefinition of fields, since this manipulation will not affect the scattering amplitudes, and of course the equations of motion as well. Hence, these considerations shall not exclude the possibility of introducing a KR-induced LV Lagrangian.
	%We note that a recent publication argued that nonbirefringence might serve as a key feature of quantum gravity, because the extended action allowing nonbirefringence meets with the effective action of type II theories~\cite{birefringence-string}. Previous research indicates that Bumblebee theory predicts birefringence while parity violation with KR field forbids this effect~\cite{birefringence-Zhu}. But since the relation with axion field is often regarded as one of the motivations for introducing KR field, take into account that \textcolor{blue}{the duality is particularly well-motivated in the context of the $E_8\times E_8$ heterotic string, and that heterotic string theories (whether $E_8\times E_8$ or $SO(32)$, the latter being dual to Type I) have no direct duality with Type II superstrings,} this shall not exclude the possibility of introducing KR field.
	
	%the duality occurs because of the compactification procedure applied on the bosonic string side in heterotic string theory
	
	%The duality between the KR field and the axion field shouldn't be astonishing.
	%And the axions were originally postulated to solve the strong CP problem in QCD. 
	Since black holes with an axion hair have long been studied, a Lorentz invariant KR-induced axion-KR solution was soon suggested in Ref.~\cite{induced-KR}. A non-minimal coupling of the KR action inspired by superstring theory also appears to admit an axion hair.

	%\textcolor{red}{The no-hair theory, especially the scalarization of black holes, are widely studied\cite{bibid}. Ref.~\cite{bibid} calculate the specific light ring during the scalarization of a black hole.} \textcolor{teal}{but there might be dilaton or other kinds of scalar fields.}
	
	%It was found that an axion hair can exist outside the event horizon of a black hole\cite{bibid}.
	
	%in which the interaction terms in the action that can cause the possible Lorentz violation are not considered, while the dynamical part of the Kalb-Ramond field, apart from its VEVs, if any, in the Lorentz violating cases, are considered.
	
	It was not until about 15 years ago that research began on a KR-field-induced spontaneous LV model~\cite{action1}. Note that the ``preferred direction'' picture no longer applies in the model discussed here~\cite{explicit-LV}. More recently, a static and spherically symmetric black hole solution was derived within this framework under the assumption of a non-dynamical KR field---implying that the field possesses only a specific VEV configuration~\cite{KR-BH}. Shortly thereafter, the same group proposed an electrically charged spherically symmetric KR black hole solution~\cite{KR-EM-BH}.
	Reference~\cite{KR-BH2} addresses the more general case of static, neutral, spherically symmetric black holes, and the authors subsequently extended this work to slowly rotating KR black holes later~\cite{KR-BH-rotate}.
	Various aspects of these black hole solutions such as the evaporation process have been extensively studied~\cite{Filho1,Filho2,Filho3,Filho4}. Ref.~\cite{FilhoBH} presents a new regular black hole solution within the framework of a non–commutative gauge theory applied to KR gravity.
	
	%However, first  the KR field minimally coupled to gravity yields to an axion hairy black hole that deforms the event horizon into a naked singularity.
	%In Ref.~\cite{action2}, ..., in which singularity is surrounded by a modified event horizon, thereby satisfying the cosmic censorship conjecture.
	
	%(the problem is whether the KR black hole are with hair? see some discussions with ZYP shixiong)
	%\textcolor{violet}{first, carrying a hair or not has nothing to do with whether the corresponding field is dynamical or not?} \textcolor{Chartreuse4}{secondly, it seems that the Lorentz violation parameter can be absorbed by the other hairs, so maybe it shouldn't be considered as a independent hair.}
	
	In this work, rather than studying birefringence, we focus on the quasinormal modes (QNMs) of such a black hole solution obtained in a spontaneous Lorentz-violating scenario. While the QNMs of test scalar, Dirac, and vector fields in this spacetime have been previously investigated~\cite{Filho2,KR-qnm}, we perform non-test gravitational and electromagnetic fields and consider their coupled system. The study of QNMs is crucial because the very existence of black holes remains an open question. None of the direct or indirect detections of ``black holes'' have been able to verify their existence. Chances are that the signals originate from other compact objects. In gravitational-wave astronomy, it is the ringdown stage in the merger process of the binary systems that tells us the final state of their merger. This stage corresponds to the perturbation problem under certain spacetime considered---including the black hole background---the unique boundary condition of which at the horizon reflects the intrinsic dissipative feature of black holes. In particular, this constitutes an eigenvalue problem, the eigenstates of which are called QNMs. Although studies on QNMs started long before the first successful detection of gravitational waves, identifying the nature of the compact object was their main motivation at the very beginning~\cite{old-review-qnm1,old-review-qnm2}. So the advent of gravitational-wave astronomy~\cite{GW150914} has renewed interest in QNMs, while of course gravitational-wave astronomy cannot confirm the observation of black holes to date, since the signal-to-noise ratio in the ringdown stage remains insufficient to extract the QNMs in current observations.

	The recent rise in interest in QNMs has also been driven by the discovery of their applications in the gauge-gravity duality, which was first proposed by Maldacena in string theory~\cite{AdS-CFT-Maldacena}. A correspondence between QNMs and the thermalization time scale in the dual conformal field theory (CFT) was first suggested and a qualitative agreement was found~\cite{first-suggested-conjuctured}. Subsequent work demonstrated a precise quantitative agreement between quasinormal frequencies (QNFs) and the poles of the retarded thermal correlation function, with exact expressions for the QNFs of various spin for the BTZ black hole~\cite{conjuctured}. A more general discussion was presented soon after~\cite{established}. General black branes have QNFs of a ``hydrodynamic'' form. This has led to a universal value $1/4\pi$ of the ratio of shear viscosity to entropy density of the dual QFT in the limit of gravity dual description~\cite{hydrodynamic mode-shear,generalized1-hydrodynamic mode-shear,generalized2-hydrodynamic mode-shear}. This key result aids research into quark-gluon plasma (QGP) behavior and other topics.
	
	%The main interest are on the CFT side to study the properties of quark-gluon plasma (such as the sheer viscosity~\cite{AdS-CFT-sheer,review-qnm-duality,review-qnm-duality1,review-qnm-duality2}), which are important in the QCD phase transition expected to exist in early universe. Meanwhile, there are researches on the application of field theory methods to QNMs as well~\cite{QNM-duality1992,QNM-duality1,}. For a more resent complete review of QNMs, see e.g.~\cite{review-qnm}
	
	Although QNFs can be analytically explained as the poles of the retarded Green's function, they must generally be solved numerically, because the independent solutions of the perturbation equation can only be written analytically in a non-compact continued fraction form~\cite{review-qnm,CF-Leaver}. Various numerical methods have been developed, among which the continued fraction method and the direct integral method are generalized to multi-field cases~\cite{Matrix-method-CF,DIferrari2007new}. The prevailing issue, however, is that in most cases when numerical precision is adequate, the frequencies of fundamental modes are identified merely by ranking their damping rate irrespective of the underlying physical state.
	
	In this paper, we study the QNMs of an electrically charged spherically symmetric KR black hole proposed in Ref.~\cite{KR-EM-BH}. We organize this paper as follows. In Sec.~\ref{BH} we briefly review the electrically charged spherically symmetric black hole solution in a spontaneously Lorentz-violating theory induced by the VEV of the KR field. A minor error which is insignificant at the background level is identified to ensure correct treatment of the problem at the perturbation level. Then, in Sec.~\ref{Perturbation} we study the perturbation problem of the solution. The odd-parity perturbation equations of both the gravitational field and the electromagnetic field are derived. We illustrate that these two equations cannot be decoupled. Our results and analysis will be shown in Sec.~\ref{QNM}, where the numerical methods and our simple method are introduced. We apply and test this method for recognizing the different categories, and use it to test our results and perform error analysis. The effects of LV on the QNMs are analyzed in the same section. Finally we give a conclusion to our work in Sec.~\ref{Conclusion}.

\section{Electrically charged spherically symmetric KR black hole}\label{BH}
    We briefly review the electrically charged spherically symmetric KR black hole proposed by K. Yang et al~\cite{KR-EM-BH}. Introducing the KR field $B_{\mu\nu}$ which has an appropriate VEV and a non-minimal coupling with gravity, the local Lorentz symmetry breaks spontaneously. The action can be written as
    \begin{equation}
    	\begin{aligned}S&=\frac{1}{2}\int \md^4x\sqrt{-g}\bigg[R-2\Lambda-\frac{1}{6}H^{\mu\nu\rho}H_{\mu\nu\rho}-{V}(B^{\mu\nu}B_{\mu\nu}\pm b^2)\\&+\xi_2B^{\rho\mu}B^\nu{}_{\mu}R_{\rho\nu}+\xi_3B^{\mu\nu}B_{\mu\nu}R\bigg]+\int \md^4x\sqrt{-g}\mathcal{L}_{\mathrm{M}}\end{aligned}
    \end{equation}
    with $H_{\mu\nu\rho}=\partial_{[\mu} B_{\nu\rho]}$ the KR field strength, $\xi_2$ and $\xi_3$ two coupling constants, and $\Lambda$ the cosmological constant. The self-interacting potential $V(x)$ is set to reach its minimum at $x=0$. The interaction between the KR field and the gravitational field follows Ref.~\cite{action2}, where the $\xi_1$ term considered in the general SME situation~\cite{action1} has already been set to zero to obtain an analytic solution.
    %The interaction between KR field and gravitational field follows Ref.~\cite{action2}, where the $\xi_1$ term considered in the general situation~\cite{action1} which is allowed in the SME~\cite{explicit-LV} can be absorbed by redefine the gravitational coupling constant in KR vacuum. It is worth mention that resent paper~\cite{KRBHLJZ} argued that this step leads to inequivalence under variation.
    %It lacks an extra term considered in the general situation~\cite{action1} which is allowed in the SME~\cite{explicit-LV} for simplicity.
    
    The KR field can be decomposed into pseudo-electric and pseudo-magnetic parts $B_{\mu\nu}=\tilde{E}_{[\mu}v_{\nu]}+\epsilon_{\mu\nu\alpha\beta}v^{\alpha}\tilde{B}^{\beta}$, where $v$ is a timelike 4-vector. It is assumed that the vacuum configuration of the KR field exhibits a pseudo-electric configuration $b_{10}=-b_{01}=\tilde{E}(r)$, which thereby results in a vanishing KR field strength. Similar to the RN case, the background electromagnetic field in a spherical background can be set to an electrostatic form $A_\mu=-\Phi(r)\delta_\mu{}^t$.
    
    The Lagrangian of matter containing a non-minimal coupling between the electromagnetic field and the KR field is
    \begin{equation}
    	\mathcal{L}_{\mathrm{M}}=-\frac{1}{2}F^{\mu\nu}F_{\mu\nu}-\eta B^{\alpha\beta}B^{\gamma\rho}F_{\alpha\beta}F_{\gamma\rho},
    \end{equation}
    where $F_{\mu\nu}=\partial_{[\mu} A_{\nu]}$ is the electromagnetic field strength, and $\eta$ is a coupling constant. The non-minimal coupling between the KR field and the electromagnetic field was proposed to support a consistent charged black hole. The three equations of motion can then be derived by varying the action with respect to the gravitational field, the electromagnetic field and the KR field as
    \begin{equation}
    	\begin{aligned}
    		R_{\mu\nu}-\frac{1}{2}g_{\mu\nu}R+\Lambda g_{\mu\nu}&=T_{\mu\nu}^{\mathrm{M}}+T_{\mu\nu}^{\mathrm{KR}},\\
    		\nabla^\nu\left(F_{\mu\nu}+2\eta B_{\mu\nu}B^{\alpha\beta}F_{\alpha\beta}\right)&=0,\\
    		\nabla^{\alpha}H_{\alpha\mu\nu}+3\xi_2 R_{\alpha[\mu}B^{\alpha}{}_{\nu]}-6V^{\prime}B_{\mu\nu}-12\eta B^{\alpha\beta}F_{\alpha\beta}F_{\mu\nu}&=0,
    	\end{aligned}
    \end{equation}
    where
    \begin{equation}\label{EM}
    	\begin{aligned}
    		T_{\mu\nu}^{\mathrm{M}}&= 2F_{\mu\alpha}F_{\nu}{}^{\alpha}-\frac{1}{2}g_{\mu\nu}F^{\alpha\beta}F_{\alpha\beta}+\eta\left(8B^{\alpha\beta}F_{\alpha\beta}B_{\left(\mu\right.}{}^{\gamma}F_{\left.\nu\right)\gamma}-g_{\mu\nu}B^{\alpha\beta}B^{\gamma\rho}F_{\alpha\beta}F_{\gamma\rho}\right), \\
    		T_{\mu\nu}^{\mathrm{KR}}&= \frac{1}{2}H_{\mu\alpha\beta}H_{\nu}{}^{\alpha\beta}-\frac{1}{12}g_{\mu\nu}H^{\alpha\beta\rho}H_{\alpha\beta\rho}+2V^{\prime}B_{\alpha\mu}B_{~\nu}^{\alpha}-g_{\mu\nu}V \\
    		&+\xi_{2}\bigg[\frac{1}{2}g_{\mu\nu}B^{\alpha\gamma}B^{\beta}{}_{\gamma}R_{\alpha\beta}-B^{\alpha}{}_{\mu}B^{\beta}{}_{\nu}R_{\alpha\beta}-B^{\alpha\beta}B_{\nu\beta}R_{\mu\alpha}-B^{\alpha\beta}B_{\mu\beta}R_{\nu\alpha} \\
    		&+\frac{1}{2}\nabla_{\alpha}\nabla_{\mu}\left(B^{\alpha\beta}B_{\nu\beta}\right)+\frac{1}{2}\nabla_{\alpha}\nabla_{\nu}\left(B^{\alpha\beta}B_{\mu\beta}\right)-\frac{1}{2}\nabla^{\alpha}\nabla_{\alpha}\left(B_{\mu}{}^{\gamma}B_{\nu\gamma}\right) \\
    		&-\frac{1}{2}g_{\mu\nu}\nabla_{\alpha}\nabla_{\beta}\left(B^{\alpha\gamma}B^{\beta}{}_{\gamma}\right)\bigg].
    	\end{aligned}
    \end{equation}
    One might wonder why there is no $\xi_3$ term in the equations. Some researchers stated that this term can be absorbed by redefining the gravitational coupling constant in KR vacuum. It is worth mentioning that recent work~\cite{KRBHLJZ} argued that this procedure leads to inequivalence under variation. Nevertheless, these equations can be viewed as obtained by setting $\xi_3=0$ for simplicity, similar to the treatment of the $\xi_1$ term.
    
    In the expression of $T^M$ given in Ref.~\cite{KR-BH}, the first term in the bracket appears to be inconsistent, likely due to an issue in handling the variation of the contraction with a symmetric tensor. This corresponding term in Ref.~\cite{KR-BH} is obviously asymmetric, whereas a symmetrized form, as presented in Eq.~\eqref{EM}, is expected from a more detailed calculation. This discrepancy does not affect the background field equations, as both forms yield identical results under the specific configurations adopted for the background gravitational field, the KR field VEV and the background electromagnetic field, since only two diagonal components, namely (0,0) and (1,1), contribute. However, the distinction may become relevant in the context of perturbations, where non-diagonal components are generally nonzero and could thus be sensitive to such differences.
    %But since we are going to study the perturbation problem to linear order, the metric are not necessarily diagonal and the nondiagonal components of the electromagnetic field may not vanish as well. That means we should use the right formulation here.

    Considering a static and spherically symmetric metric
    \begin{equation}
    	\md s^2=-F(r)\md t^2+G(r)\md r^2+r^2\md\theta^2+r^2\sin^2\theta\md\phi^2,
    \end{equation}
    under which we have $\tilde{E}(r)=\pm\sqrt{\frac{b^2F(r)G(r)}{2}}$ (the sign does not matter since there are only quadratic terms of the KR field in the equations of motion), a solution was found in Ref.~\cite{KR-BH}. The equations of motion imply $G(r)=F^{-1}(r)$. For the case of $\Lambda=0$, the result is~\cite{KR-BH}
    \begin{equation}
    	\begin{aligned}
    		&\Phi(r)=\frac{Q}{\left(1-\ell\right)r},\\&F(r)=\frac{1}{1-\ell}-\frac{2M}{r}+\frac{Q^{2}}{\left(1-\ell\right)^{2}r^{2}},
    	\end{aligned}
    \end{equation}
    where $l\equiv\xi_2b^2/2$ is the LV parameter. A similar solution has also been suggested within the framework of Bumblebee gravity~\cite{bumblebeeBHLJZ}. Solving $F(r)=0$, we obtain the horizon radii
    \begin{equation}
    	r_\pm=(1-\ell)\left(M\pm\sqrt{M^2-\frac{Q^2}{(1-\ell)^3}}\right),
    \end{equation}
    which give a constraint on the parameters $Q^{2}/M^{2}\leq(1-\ell)^{3}$ for a black hole solution. Note that this constraint also implies that $\ell\leq1$. In the following part of this paper, we will refer to this solution as a charged KR black hole.

\section{Perturbation problem}\label{Perturbation}
\subsection{Perturbation equations}
    We follow the general process proposed by Regge and Wheeler~\cite{RW} to derive the perturbation equations. 
    %This procedure suits all the cases with a spherical background.
    The perturbed metric and electrmagnetic field can be written as %(\textcolor{blue}{why not consider KR perturbation?}
    \begin{equation}
    	\begin{array}{rcl}g_{\mu\nu}&=&\bar{g}_{\mu\nu}+h_{\mu\nu},\\A_\mu&=&\bar{A}_\mu+a_\mu,\end{array}
    \end{equation}
    where the quantities with ``bar'' refer to the corresponding background field, and the perturbation fields are referred by $h_{\mu\nu}$ and $a_\mu$. We first decompose the perturbations into scalar, vector and tensor parts according to SO(3) symmetry:
    \begin{equation}
    	\NiceMatrixOptions{code-for-first-row = \color{blue}}
    	a_\mu=
    	\begin{bNiceArray}{cc|cc}[first-row,first-col]
    		&t  &r  &\;\theta &\;\;\phi\;\; \\
    		&s1 &s2 &\Block{1-2}{v1} & \\
    	\end{bNiceArray},
        \qquad\qquad
        h_{\mu\nu}=
    	\NiceMatrixOptions{code-for-first-row = \color{blue},
    		code-for-first-col = \color{blue}}
    	\begin{bNiceArray}{cc|cc}[first-row,first-col]
    		&t  &r  &\theta &\;\;\phi\;\; \\
    		t &s1 &s2 &\Block{1-2}{v1} & \\
    		r &s2 &s3 &\Block{1-2}{v2} & \\
    		\hline
    		\theta &\Block{2-1}{v1^\top} &\Block{2-1}{v2^\top} &\Block{2-2}{t} & \\
    		\phi & & & & \\
    	\end{bNiceArray}.
    \end{equation}
    
    By using the spherical harmonic bases for SO(3) scalars
    \begin{equation}
    	\phi_{L}{}^{M}\propto Y_{L}{}^{M}(\theta,\varphi),\quad\mathrm{even},
    \end{equation}
    vectors
    \begin{equation}
    	\begin{aligned}
    		\psi_{L}{}^{M}{}_{a}&\propto Y_{L}{}^{M}{}_{;a}(\theta,\varphi),&&\mathrm{even},\\
    		\phi_{L}{}^{M}{}_{a}&\propto \epsilon_{a}{}^{b}Y_{L}{}^{M}{}_{;b}(\theta,\varphi),&&\mathrm{odd},
    	\end{aligned}
    \end{equation}
    and tensors
    \begin{equation}
    	\begin{aligned}
    		\psi_{L}{}^{M}{}_{ab}&\propto Y_{L}{}^{M}{}_{;ab}(\theta,\varphi),&&\mathrm{even},\\
    		\phi_{L}{}^{M}{}_{ab}&\propto \gamma_{ab}Y_{L}{}^{M}(\theta,\varphi),&&\mathrm{even},\\
    		\chi_{L}{}^{M}{}_{ab}&\propto \frac{1}{2}[\epsilon_{a}{}^{c}\psi_{L}{}^{M}{}_{cb}(\theta,\varphi)+\epsilon_{b}{}^{c}\psi_{L}{}^{M}{}_{ca}(\theta,\varphi)],&&\mathrm{odd},
    	\end{aligned}
    \end{equation}
    we divide the perturbations into odd and even parity parts, where the lowercase Latin letter indices $a$, $b$ and $c$ refer to $\theta$ and $\phi$ and ``odd/even parity'' parts acquire a factor $(-)^{L+1}$/$(-)^L$ under parity transformation. $\boldsymbol{\epsilon}$ is the two-dimensional volume element (distinguish from the total antisymmetric tensor density $\epsilon$), and $\gamma_{ab}=\bar{g}_{ab}/r^2$ is the metric tensor on the sphere. We will only consider the $M=0$ cases because of spherical symmetry, since the radial equations are independent of $M$ after separating variables. In the following context, the index $M$ will be omitted to prevent notational conflict with the black hole mass $M$.
    
    Thanks to gauge invariance, the form of gravitational perturbations can be vastly simplified. Given a gauge transformation for odd and even gravitational perturbations respectively of the form $\xi_{\mu;\nu}+\xi_{\nu;\mu}$, where the ``coordinate transformation'' covector $\xi^\mu$ can always be written in the general form
    \begin{equation}
    	\begin{aligned}
    		\xi^\mu_\mathrm{odd}&=
    		\left[\begin{array}{cc|c}
    			0 &0 &\Lambda(t,r)\epsilon^{ab}Y_{L}{}_{;b}(\theta,\varphi)
    		\end{array}\right],\\
    	    \xi^\mu_\mathrm{even}&=
    	    \left[\begin{array}{cc|c}
    	    	\Theta_{0}(t,r)Y_{L}(\theta,\varphi) &\Theta_{1}(t,r)Y_{L}{}(\theta,\varphi) &\Theta(t,r)\gamma^{ab}Y_{L}{}_{;b}(\theta,\varphi)
    	    \end{array}\right],
    	\end{aligned}
    \end{equation}
    we can always reduce the perturbation into a simple form
    \begin{equation}
    	\begin{aligned}
    		h_{\mu\nu}^\mathrm{odd}&=\begin{bmatrix}0 & 0 & 0 & h_0(r)\\ 0 & 0 & 0 & h_1(r)\\ 0 & 0 & 0 & 0\\ \text{Sym} & \text{Sym} & 0 & 0\end{bmatrix}\times\exp(-i\omega t)\sin\theta~\partial_\theta P_{L}(\cos\theta),\\
    		h_{\mu\nu}^\mathrm{even}&=
    		\begin{bmatrix}
    			F(r)H_0(r)&H_1(r)&0&0\\
    			\text{Sym}&F^{-1}(r)H_2(r)&0&0\\
    			0&0&r^2K(r)&0\\
    			0&0&0&r^2K(r)\sin^2\theta
    		\end{bmatrix}
    	    \times\exp(-i\omega t)P_L(\cos\theta)
    	\end{aligned}
    \end{equation}
    by solving the suitable coefficients $\Lambda,\Theta_0,\Theta_1$, and $\Theta$.
    
    The important thing here is that the gauge condition can be chosen for odd and even parity respectively. Although the gravitational gauge transformations are usually considered as coordinate transformations, we are here in the perturbation theory treating them as real gauge transformations without changing the coordinates~\cite{Carroll}. And since we are not taking them as coordinate transformations, there is no need to worry about the perturbations of electromagnetic field changing under these gauge transformations.
    
    As mentioned before, this gauge-fixing procedure always work under a spherical background. One might check it and find that fixing the gauge is irrelevant to the specific choice of background metric as long as it has a spherical configuration.
    
    As for the electromagnetic perturbation, it is just a covariant vector, so we directly write it into a simple form with our basic vectors
    \iffalse
    \begin{equation}
    	\begin{split}
    		a^\mathrm{odd}_\mu&=
    		\left[\begin{array}{cc|c}
    			0 &0 &h_v(t,r)\epsilon_{\mu}{}^{\nu}(\partial/\partial x^{\nu})Y_{L}{}^{M}(\theta,\varphi)
    		\end{array}\right]\\
    	    &\xlongequal{M=0}
    	    \left[\begin{array}{cccc}
    	    	0 &0 &0 &h_v(r)
    	    \end{array}\right]\times\exp(-i\omega t)(\sin(\theta)\partial/\partial\theta)P_L(\cos\theta),\\
        \end{split}
    \end{equation}
    \begin{equation}
    	\begin{split}
    		a^\mathrm{even}_\mu&=
    		\left[\begin{array}{cc|c}
    			H_{v0}(t,r)Y_{L}{}^{M}(\theta,\varphi) &H_{v1}(t,r)Y_{L}{}^{M}(\theta,\varphi) &H_v\left(t,r\right)(\partial/\partial x^{\mu})Y_{L}{}^{M}(\theta,\varphi)
    		\end{array}\right]\\
    	    &\xlongequal{M=0}
    	    \left[\begin{array}{cccc}
    	    	H_{v0}(r) &H_{v1}(r) &H_v(r)\partial/\partial\theta &0
    	    \end{array}\right]\times\exp(-i\omega t)P_L(\cos\theta).
    	\end{split}
    \end{equation}
    \fi
    \begin{equation}
    	\begin{aligned}
    		a^\mathrm{odd}_\mu&=\left[\begin{array}{cccc}
    			0 &0 &0 &h_v(r)
    		\end{array}\right]\times\exp(-i\omega t)\sin\theta~\partial_\theta P_L(\cos\theta),\\
    		a^\mathrm{even}_\mu&=\left[\begin{array}{cccc}
    			H_{v0}(r) &H_{v1}(r) &H_v(r)\partial_\theta &0
    		\end{array}\right]\times\exp(-i\omega t)P_L(\cos\theta).
    	\end{aligned}
    \end{equation}
    %加上电磁规范的问题：说一下研究奇宇称微扰，规范不影响
    We focus on the odd parity perturbation, and follow the gauge choice in Ref.~\cite{gauge-Zerilli}, such that the odd parity perturbation does not change after gauge fixing.
    
    Substituting the perturbed fields into the Einstein's equations and the Maxwell's equations, expanding them to linear order, we can get the perturbation equations. Thanks again to the spherical symmetry of the system, the odd and even parity parts of the equations can be decoupled, and the results are just equal to those when considering only odd or even parity perturbations. We will focus on the odd parity parts. Due to the symmetry of indices of the Einstein's equations, just four equations are left. Obviously, with merely three variables, only three equations are independent. Furthermore, discovering that one equation acts as a constraint leaves precisely two equations. Transforming into tortoise coordinate $r_\ast$ defined as $\md r_\ast=\frac{\md r}{F(r)}$, and applying a variable substitution to the two variables left,
    \begin{equation}
    	\psi_\mathrm{g}\equiv \frac{F(r)}{\omega r}h_1\qquad\text{and}\qquad\psi_\mathrm{e}\equiv h_v,
    \end{equation}
    the equations can be further simplified
    \begin{equation}\label{pert-eq}
    	\begin{aligned}
    		\frac{\md^{2}\psi_{\mathrm{g}}}{\md r_{*}^{2}}+(\omega^{2}-V_\mathrm{gg})\psi_{\mathrm{g}}-V_\mathrm{ge}\psi_{\mathrm{e}}=0,\\
    		\frac{\md^{2}\psi_{\mathrm{e}}}{\md r_{*}^{2}}+(\omega^{2}-V_\mathrm{ee})\psi_{\mathrm{e}}-V_\mathrm{eg}\psi_{\mathrm{g}}=0,
    	\end{aligned}
    \end{equation}
    where the effective potentials are given by
    \begin{equation}
    	\begin{aligned}
    		V_\mathrm{gg}=&\left[\frac{(4-6\ell)Q^2+(\ell-1)^2r\left((8\ell-6)M+(L(L+1)-(L(L+1)-2)\ell)r\right)}{(\ell-1)^2(2\ell-1)r^4}-\frac{2\ell F(r)}{(2\ell-1)r^2}\right]F(r),\\
    		V_\mathrm{ge}=&\frac{4iQF(r)}{(2\ell-1)r^3},\qquad V_\mathrm{ee}=\frac{(4Q^2+(1-2\ell)L(L+1)r^2)F(r)}{(2\ell-1)r^4},\qquad
    		V_\mathrm{eg}=\frac{i(\ell-1)(L(L+1)-2)QF(r)}{(2\ell-1)r^3}.
    	\end{aligned}
    \end{equation}
    These equations do recover to the ones under a RN background when the LV parameter $l\rightarrow0$, which is consistent with our expectation.

\subsection{Decoupling of the equations}\label{decouple}
    It can be shown that these two equations cannot be decoupled. We generally follow the decouple process in Refs.~\cite{decouple-odd,decouple-even}. The basic idea is to make use of diagonalization of matrices to eliminate the coupled terms in the equation set.
    
    Equation~\eqref{pert-eq} can be easily written into a matrix equation form
    \begin{equation}\label{coupled-equation}
    	\left(\partial_{r_\ast}^2+\omega^2\right)
    	\left(\begin{array}{c}
    		\psi_{\mathrm{g}}\\
    		\psi_{\mathrm{e}}
    	\end{array}\right)
        =\mathbf{V}_{eff}
        \left(\begin{array}{c}
        	\psi_{\mathrm{g}}\\
        	\psi_{\mathrm{e}}
        \end{array}\right),
    \end{equation}
    where
    \begin{equation}
    	\mathbf{V}_\mathrm{eff}=
    	\left(\begin{array}{cc}
    		V_\mathrm{gg} &V_\mathrm{ge}\\
    		V_\mathrm{eg} &V_\mathrm{ee}
    	\end{array}\right).
    \end{equation}
    Our main purpose is to decompose the effective potential matrix $\mathbf{V}_\mathrm{eff}$ into the following form
    \begin{equation}\label{eff-p-dcp}
    	\mathbf{V}_\mathrm{eff}\equiv\sum_{i}f_i(r)\mathbf{C}_i,
    \end{equation}
    where $f_i(r)$ are some functions of $r$, $\mathbf{C}_i$ are radial-independent matrices, which can be diagonalized by a radial-independent transformation matrix $\mathbf{S}$. For example, if there is just one term in the summation on the r.h.s. of Eq.~\eqref{eff-p-dcp}, then the effective potential can be diagonalized:
    \begin{equation}\label{diag}
    	\begin{split}
    		\left(\partial_{r_\ast}^2+\omega^2\right)
    		\left(\begin{array}{c}
    			\psi_{\mathrm{g}}\\
    			\psi_{\mathrm{e}}
    		\end{array}\right)
    		&=f_1(r)\mathbf{S}^{-1}\mathbf{S}\mathbf{C}_1\mathbf{S}^{-1}\mathbf{S}
    		\left(\begin{array}{c}
    			\psi_{\mathrm{g}}\\
    			\psi_{\mathrm{e}}
    		\end{array}\right),\\
    	    \Rightarrow
    	    \left(\partial_{r_\ast}^2+\omega^2\right)\left[\mathbf{S}
    	    \left(\begin{array}{c}
    	    	\psi_{\mathrm{g}}\\
    	    	\psi_{\mathrm{e}}
    	    \end{array}\right)\right]
            &=f_1(r)\left[\mathbf{S}\mathbf{C}_1\mathbf{S}^{-1}\right]\left[\mathbf{S}
            \left(\begin{array}{c}
            	\psi_{\mathrm{g}}\\
            	\psi_{\mathrm{e}}
            \end{array}\right)\right].\\
    	\end{split}
    \end{equation}
    Since $\mathbf{S}$ is independent of the radial coordinate, it commutes with the operator $\left(\partial_{r_\ast}^2+\omega^2\right)$ as shown on the l.h.s. of Eq.~\eqref{diag}.
    In general situations, if and only if all the $\mathbf{C}_i$'s can be diagonalized simultaneously, we can decouple Eq.~\eqref{coupled-equation}
    \begin{equation}
    	\left(\partial_{r_\ast}^2+\omega^2\right)
    	\left(\begin{array}{c}
    		\psi_{\mathrm{g}}\\
    		\psi_{\mathrm{e}}
    	\end{array}\right)^\prime
    	=\sum_{i}f_i(r)\mathbf{C}_i^{\mathbf{diag}}
    	\left(\begin{array}{c}
    		\psi_{\mathrm{g}}\\
    		\psi_{\mathrm{e}}
    	\end{array}\right)^\prime.
    \end{equation}
    In our case of the charged KR black hole
    \begin{equation}
    	\left(\partial_{r_\ast}^2+\omega^2\right)
    	\left(\begin{array}{c}
    		\psi_{\mathrm{g}}\\
    		\psi_{\mathrm{e}}
    	\end{array}\right)
    	=F(r)\left(\mathbf{C}_2r^{-2}+\mathbf{C}_3r^{-3}+\mathbf{C}_4r^{-4}\right)
    	\left(\begin{array}{c}
    		\psi_{\mathrm{g}}\\
    		\psi_{\mathrm{e}}
    	\end{array}\right),
    \end{equation}
    where
    \begin{equation}
    	\begin{split}
    		\mathbf{C}_2&=\left(\begin{array}{cc}
    			-\frac{(1-2l)L(L+1)+l^2(L^2+L-2)}{(l-1)(2l-1)} &0\\
    			0 &-L(L+1)
    		\end{array}\right),\\
    	    \mathbf{C}_3&=\left(\begin{array}{cc}
    	    	6M &\frac{4iQ}{2l-1}\\
    	    	\frac{i(l-1)(L^2+L-2)Q}{2l-1} &0
    	    \end{array}\right),\qquad
    	    \mathbf{C}_4=\left(\begin{array}{cc}
    	    	\frac{(4-6l)Q^2}{(l-1)^2(2l-1)} &0\\
    	    	0 &\frac{4Q^2}{2l-1}
    	    \end{array}\right).
    	\end{split}
    \end{equation}
    The simultaneously diagonalization is impossible, since $\mathbf{C}_2$ and $\mathbf{C}_4$ are already diagonal but not identity matrices, while $\mathbf{C}_3$ is not diagonal. So these equations can not be decoupled, and we need to apply the numerical methods capable of finding the eigenvalues of a coupled equation set.

\section{Eigenvalue problem}\label{QNM}
\subsection{Numerical method}\label{CF}
\subsubsection{Matrix-valued continued fraction method}
    Due to the need to solve coupled differential equations, we choose the continued fraction method, or the Frobenius method, proposed by E. W. Leaver~\cite{CF-Leaver}. Here we take a brief sight of a general introduction to this method following~\cite{CF-others-Zhidenko}. Given an ordinary differential equation
    \begin{equation}\label{ODE}
    	\left(\frac{\md^2}{\md r^2}+p(r)\frac{\md}{\md r}+q(r)\right)R(r)=0,
    \end{equation}
    in an asymptotically flat spacetime, a series solution can be written in the following form
    \begin{equation}
    	R(r)=e^{i\Omega r}(r-r_{0})^{\sigma}\left(\frac{r-r_{+}}{r-r_{0}}\right)^{-i\beta}\sum_{k=0}^{\infty}b_{k}\left(\frac{r-r_{+}}{r-r_{0}}\right)^{k},
    \end{equation}
    where $\Omega$, $\sigma$ and $\beta$ are determined by boundary conditions, $r_{+}$ is the event horizon, and $r_{0}$ an arbitrary parameter smaller than $r_{+}$. This is the solution of Eq.~\eqref{ODE} if and only if the series is convergent. Substituting the series solution into the equation, one can obtain an $N$-term recurrence relation for the coefficients $b_i$
    \begin{equation}\label{series}
    	\sum_{j=0}^{\min(N-1,i)}c_{j,i}^{(N)}(\omega)b_{i-j}=0,\quad i>0.
    \end{equation}
    One might think Eq.~\eqref{series} is obtained from a power series, but since we should change all the variable $r$ to $x=\left(\frac{r-r_{+}}{r-r_{0}}\right)$, when we substitute $R(r)$ into the equation, we have to multiply by a function of $x$ to pick up terms of $x^n$ and obtain the recurrence equations. This process may be thought of as a change of function bases.
    %\textcolor{blue}{We are not sure whether the orthogonality and completeness of the bases have been proven, but it is a problem for the mathematicians. At first glance we can prove it by just redefining the inner product. }
    But we will continue using ``$x^n$ term'' for simplicity.
    
    Usually, we shall decrease the number of terms in the recurrence relation to three by use of Gaussian eliminations
    \begin{subequations}
    	\begin{align}
    		&c_{0,i}^{(3)}b_i+c_{1,i}^{(3)}b_{i-1}+c_{2,i}^{(3)}b_{i-2}=0,\quad i>1\label{gauss:sub1}\\
    		&c_{0,1}^{(3)}b_1+c_{1,1}^{(3)}b_0=0.\label{gauss:sub2}
    	\end{align}
    \end{subequations}
    Equation~\eqref{gauss:sub2} directly shows
    \begin{equation}\label{rhs1}
    	\frac{b_1}{b_0}=-\frac{c_{1,1}^{(3)}}{c_{0,1}^{(3)}},
    \end{equation}
    while Eq.~\eqref{gauss:sub1} gives a continued fraction form of $b_1/b_0$
    \begin{equation}\label{rhs2}
    	\frac{b_1}{b_0}=-\frac{c_{2,2}^{(3)}}{c_{1,2}^{(3)}-}\frac{c_{0,2}^{(3)}c_{2,3}^{(3)}}{c_{1,3}^{(3)}-}\frac{c_{0,3}^{(3)}c_{2,4}^{(3)}}{c_{1,4}^{(3)}-}\ldots\ .
    \end{equation}
    So the equation has solution only if the r.h.s.'s of Eq.~\eqref{rhs1} and Eq.~\eqref{rhs2} are equal. And if we set the boundary solutions at both sides as eigenstates for the same eigenvalue $\omega$, this gives the eigenvalue equation, which can be numerically solved.
    
    %\textcolor{blue}{the question is does $\omega$ of the two equations/variables have to be the same one? According to the two references~\cite{bibid}, seem to be ``yes''. But if so, how can we restore the RN/Schw scenarios?}
    
    However, here we do not follow the usual procedure of finding the eigenvalue equation. Instead, we obtain the QNFs, i.e., the eigenvalues of the equation, by solving the secular equation of the set of recurrence relations~\cite{Matrix-method-Guo}. This generalization to equation sets was first proposed in Ref.~\cite{Matrix-method-CF}. In the case of two coupled equations, Eq.~\eqref{series} can be written as a matrix equation
    \begin{equation}\label{recurrence-eq}
    	\sum_{j=0}^{\min(N-1,i)}\mathbf{C}_{j,i}^{(N)}(\omega)\mathbf{b}_{i-j}=0,\quad i>0,
    \end{equation}
    where $\mathbf{C}_{j,i}^{(N)}(\omega)$ are the coefficient two by two matrices of the $i$th recurrence equation, and $\mathbf{b}_{i-j}$ are two by one vectors. Here, the phrase ``the $i$th'' represents the coefficient equation of $x^i$, acquired by substituting the $x^j$ term in the series solutions into the equation set and taking the coefficients of $x^{i}$. The four components of $\mathbf{C}_{j,i}^{(N)}(\omega)$ correspond to the contribution of the two components of $\mathbf{b}_{i-j}$ for the two equations respectively. These recurrence equations can be organized into a block matrix form, we refer to the one in Ref.~\cite{Matrix-method-Guo}, while making a little change of notation for clarification. Theoretically, the order index $i$ should extend to infinity, but in practical calculations, we must truncate the recurrence system at a finite maximum order $i_{\max}$. The time of calculation increases rapidly when we raise the cutoff dimension. Unfortunately, unlike in Schwarzschild and RN limit, the continued fraction method does not perform sufficiently well in the case here under such low cutoff dimension. We will turn to the direct integration method.

\subsubsection{Matrix-valued direct integration method}
    The direct integration method is a very straightforward method to solve the equations of QNM problem. The initial approach involved calculating the Wronskian of the solutions integrated from both sides. However, this procedure was said to be plagued by numerical instability~\cite{DIdetweiler1975variational}. So in Ref.~\cite{DIchandrasekhar1975quasi}, a variable substitution was done, which changes the problem into solving a Riccati equation, and looses our limitation of choosing the starting point of integration. A series expansion at both sides is then made because of the approximation. The numerical method is then as described to vanish the difference of the two solutions (of a new variable $\phi$) integrated from both sides. As shown in Ref.~\cite{DIchandrasekhar1975quasi}, the numerical instability is just obscured instead of eliminated, so the method is only valid for those frequencies having a smaller absolute value of its imaginary part than its real part.
    
    A development of the method was made when searching for the QNMs of stars~\cite{DIchandrasekhar1991a}. The authors directly integrate the Zerilli equation forward from the surface of the star to determine its asymptotic behavior in~\cite{DIchandrasekhar1975quasi}. However, instead just vanishing the reflexion coefficient, in that paper and many following articles, they actually use another method called the Breit-Wigner formula~\cite{DIthorne1969}, which only focus on solving the real solutions for real frequencies, and finding the real and imaginary parts of the eigenfrequencies by fitting the quadratic behavior of its flux near the minimum points. This alternative method is valid only under the Breit-Wigner assumption $\omega_I\ll\omega_R$. Reference~\cite{DIchandrasekhar1991b} proved the equivalence of these two methods, for the real parts of the frequencies, at least. The equivalence of finding the imaginary parts can only be verified numerically.
    
    Both methods, which can be referred to as the direct integration method, have been generalized respectively into the cases of equation set~\cite{DIferrari2007new}. For a complete review, see Refs.~\cite{DIrosa2012massive,DIpani2012perturbations,DIpani2013advanced}. Here we briefly go over the procedure.
    
    For a free wave equation, the solution can be decoupled into a superposition of ingoing and outgoing plane waves. So at the boundary of our system, since the effective potentials tend to zero as $r_*\rightarrow\pm\infty$, the asymptotic solution can be expand into the formula
    \begin{equation}
    	\mathbf{u}\sim \mathbf{B}e^{-\omega r_*}+\mathbf{C}e^{\omega r_*},
    \end{equation}
    where $\mathbf{u}$, $\mathbf{B}$ and $\mathbf{C}$ are vectors. At the same time, we need to cut the boundary at both ends, so the boundary condition is a little deviated from plane wave. We write the deviation as a series expansion to the $n$th order
    %At the same time, we need to make a series expansion to the $n$th order around the plane wave at horizon and infinity as we have said before, since a cut off of the limit of integration must be made. So in practice we take the ingoing boundary condition at horizon as
    \begin{equation}
    	\mathbf{u}_\mathrm{(h)}\sim \sum_{i=0}^n{\mathbf{b}_\mathrm{(h)}}_i(r-r_+)^i\lim_{r\rightarrow r_+}e^{-\omega r_*}.
    \end{equation}
    Substituting this formula into the perturbation equations, a recurrence equation between the expansion coefficients can be obtained. Then all the coefficients can be written as combination of the zero-order coefficients ${\mathbf{b}_{(h)}}_0$
    \begin{equation}
    	\mathbf{u}_\mathrm{(h)}\sim \sum_{i}{\mathbf{m}_\mathrm{(h)}}_i{\mathbf{b}_\mathrm{(h)}}(r-r_+)^i\lim_{r\rightarrow r_+}e^{-\omega r_*}
    	\equiv\mathbf{M}_\mathrm{(h)}(\omega,r){\mathbf{b}_\mathrm{(h)}},
    \end{equation}
    where we have dropped the subscribe ``0'' for simplicity.
    The matrix $\mathbf{M}_\mathrm{(h)}(\omega,r)$ remains unchanged as long as the cutoff order $n$ does not change. So ${\mathbf{b}_\mathrm{(h)}}$ determines the boundary condition near the horizon completely. Choosing a set of orthonormal base of ${\mathbf{b}_\mathrm{(h)}}$ as ${\mathbf{b}^{(1,0)}_\mathrm{(h)}}=\left(\begin{array}{c}
    	1\\0
    \end{array}\right)$ and ${\mathbf{b}^{(0,1)}_\mathrm{(h)}}=\left(\begin{array}{c}
    0\\1
    \end{array}\right)$, any solution can be written as
    \begin{equation}
    	\mathbf{b}^{(a,b)}_\mathrm{(h)}
    	=\left[\mathbf{b}^{(1,0)}_\mathrm{(h)}\quad\mathbf{b}^{(0,1)}_\mathrm{(h)}\right]\left[\begin{array}{c}
    		a\\
    		b
    	\end{array}\right].
    \end{equation}
    Integrating $\mathbf{u}_\mathrm{(h)}$ outward, we find the corresponding values of the solutions at infinity because of the linearity of integration
    \begin{equation}
    	\mathbf{Y}^{(a,b)}
    	=\left[\mathbf{Y}^{(1,0)}\quad\mathbf{Y}^{(0,1)}\right]\left[\begin{array}{c}
    		a\\
    		b
    	\end{array}\right].
    \end{equation}
    
    A similar expansion is to be made for solutions at infinity 
    \begin{equation}\label{expand-inf}
    	\mathbf{u}_\mathrm{(inf)}\sim \sum_{i=0}^m{\mathbf{b}_\mathrm{(inf)}}_ir^{-i}\lim_{r\rightarrow \infty}e^{-\omega r_*}
    	+\sum_{i=0}^m{\mathbf{c}_\mathrm{(inf)}}_ir^{-i}\lim_{r\rightarrow \infty}e^{\omega r_*}.
    \end{equation}
    The point is that we want to test whether the solution we just obtained through integration satisfies our boundary condition at infinity. And generally we have
    \begin{equation}
    	\left[\begin{array}{c}
    		\mathbf{Y}^{(a,b)}\\
    		\partial_r\mathbf{Y}^{(a,b)}
    	\end{array}\right]
    	\equiv{\mathbf{M}_\mathrm{(inf)}}(\omega,r)
    	\left[\begin{array}{c}
    		\mathbf{B}^{(a,b)}\\
    		\mathbf{C}^{(a,b)}
    	\end{array}\right]
    \end{equation}
    %is this actually Wronskian? No
    from Eq.~\eqref{expand-inf}, where ${\mathbf{b}_\mathrm{(inf)}}_0\equiv \mathbf{B}$ and ${\mathbf{c}_\mathrm{(inf)}}_0\equiv \mathbf{C}$. We can express inversely the zero-order coefficients in terms of $\mathbf{Y}$ and get $\mathbf{B}^{(1,0)}$, $\mathbf{B}^{(0,1)}$, $\mathbf{C}^{(1,0)}$ and $\mathbf{C}^{(0,1)}$ respectively. Since all the relations are linear, it is easy to prove that
    \begin{equation}\label{eq-B}
    	\mathbf{A}^{(a,b)}
    	=\left[\mathbf{A}^{(1,0)}\quad\mathbf{A}^{(0,1)}\right]\left[\begin{array}{c}
    		a\\
    		b
    	\end{array}\right],
    \end{equation}
    where $\mathbf{A}$ refers to $\mathbf{B}$ or $\mathbf{C}$. Finally, we arrive at the conclusion that $\omega$ is a QNF only if
    \begin{equation}
    	\det\left[\mathbf{B}^{(1,0)}\quad\mathbf{B}^{(0,1)}\right]=0.
    \end{equation}
\subsubsection{A method of recognizing the eigenvalues}\label{recog}
    We develope a method of recognizing the eigenvalues of coupled equations that cannot be decoupled and estimating the degree of coupling.
    
    Strictly speaking, the eigenvalues and eigensolutions of a set of coupled equations that cannot be decoupled can only be explained as common eigenvalues and eigensolutions of the equation set. The eigenvectors do not follow a specific law. On the contrary, for those that can be decoupled, e.g. in the RN case, under the decoupled form, it is easy to distinguish the difference of the gravity-eigenfrequencies and the electromagnetic-eigenfrequencies, since most of the eigenstate vectors are just the base vectors. Of course the set of eigenfrequencies in the coupled form is the union set of the ones in the two decoupled equations, which means it may contain degenerate modes. However, the probability should be fairly low, which means we barely need to take account of them, especially when searching for fundamental modes.
    
    If the equations are still coupled, the eigenvector would be transformed by a transformation matrix, just as shown in subsection~\ref{decouple}. This might prevent us from recognizing whether the equations can be decoupled by just looking at the eigenstates. Looking at the fraction of two components of the different eigenvectors under the same coefficients may solve the problem. But the real question is: whether it is really important to decouple the equations?
    
    We have provided our proof of the reason why our perturbation equations cannot be decoupled in~\ref{decouple}. Some may doubt the non-decouplability, but the frequencies are truly the right result, since our numerical methods are reliable. It is just the explanation being different, as said above, so the problem of decoupling really does not matter.
    
    Despite all the above, for the specific problem we are dealing with, the modification caused by LV should be sufficiently small. So even if the equations cannot be decoupled, the degree of non-decouplability should be rather weak. And note that our perturbation equations degenerate to the coupled ones in RN case. Then it's reasonable to infer that the eigenstates and eigenfrequencies obey a rule with just a small deviation from the RN case. So we can still distinguish the spectral slightly deviated from the RN gravity or electromagnetic spectral by its eigenvectors $\mathbf{C}$, or $(a,b)$. Obviously, $(a,b)$ should be equal to $\mathbf{C}^{(a,b)}$ up to a scalar factor at least for decouplable equation sets, considering the decoupled forms. And the numerical results under RN limit support it, which has as well tested the effectiveness of our method. For instance, for the fundamental electromagnetic mode when $l=0.01$, $\frac{C_1'}{C_2'}=4.74314-27.9028i$, while $\frac{a}{b}=5.29009-27.9246i$. We have proven that the linear combination coefficients of $\mathbf{C}^{(a,b)}$ are the same as the ones of $\mathbf{B}^{(a,b)}$. The only job left is to solve the equation
    \begin{equation}
    	\left[\mathbf{B}^{(1,0)}\quad\mathbf{B}^{(0,1)}\right]\left[\begin{array}{c}
    		a\\
    		b
    	\end{array}\right]=0.
    \end{equation}
    The problem is that the coefficient determinant of our equation merely numerically, rather than strictly, tends to zero. To find the numerical solution, just take either of the two simultaneous equations, and solve the relation of $a$ and $b$.
    
    Remember that under the undecoupled form the eigenstates are transformed by a matrix, so the ratio would not be $1/0=\infty$ or $0/1=0$. They are completely dependent on the matrix so that we cannot distinguish the discriminant by just looking at the ratio itself. What we can know about is whether the ratio clearly separate into two categories, and we can distinguish them by other less indirect clues as shown in the following context. And since the eigenvector and transformation matrix are complex, because of both analytical and numerical reasons, it is reasonable that the ratio of the norm of two components of the eigenvalues would be adequate for recognization as long as this quantity separate clearly into two categories as said before. Take the example of eigenstate $(1,0)$:
    \begin{equation}
    	\left[\begin{array}{c}
    		C_1'\\
    		C_2'
    	\end{array}\right]=
    	\left[\begin{array}{cc}
    		S_1 &S_2\\
    		S_3 &S_4
    	\end{array}\right]\left[\begin{array}{c}
    		K\\
    		0
    	\end{array}\right]=
    	ke^{i\lambda}\left[\begin{array}{c}
    		s_1e^{i\theta}\\
    		s_3e^{i\phi}
    	\end{array}
    	\right],
    \end{equation}
    \begin{equation}
    	\frac{C_1'}{C_2'}=\frac{c_1'}{c_2'}e^{i(\theta-\phi)},\qquad
    	\frac{|C_1'|}{|C_2'|}=\frac{c_1'}{c_2'}.
    \end{equation}
    
\subsection{Results and analysis}
    As shown in~\ref{CF}, we use Frobenius method to solve the eigenvalues. The series solution is expressed as follows
    \begin{equation}
    	\begin{aligned}
    		\psi_{\mathrm{g}}(r)=&e^{i(1-\ell)\omega(r-r_+)}(r-r_+)^{-\frac{ir_+^2(1-\ell)\omega}{r_+-r_-}}(r-r_++1)^{\frac{ir_+^2(1-\ell)\omega}{r_+-r_-}+i(r_++r_-)(1-\ell)\omega}\times\sum_na_n^\mathrm{g}(\frac{r-r_+}{r-r_-})^n,\\
    		\psi_{\mathrm{e}}(r)=&e^{i(1-\ell)\omega(r-r_+)}(r-r_+)^{-\frac{ir_+^2(1-\ell)\omega}{r_+-r_-}}(r-r_++1)^{\frac{ir_+^2(1-\ell)\omega}{r_+-r_-}+i(r_++r_-)(1-\ell)\omega}\times\sum_na_n^\mathrm{e}(\frac{r-r_+}{r-r_-})^n.
    	\end{aligned}
    \end{equation}
    The fundamental QNFs for $L=2$ with different values of $Q$'s in the RN limit, i.e., $l=0$ are shown in Tables~\ref{CF1} and~\ref{CF2}. Cutting off the dimension of coefficient matrix at 40, we find that the fundamental QNFs of the charged KR black hole in the Schwarzschild limit coincides perfectly well with the ones of the Schwarzschild black hole. The fundamental QNFs of the charged KR black hole in the RN limit has been calculated to test the validity of our code as well. We do not take the same cutoff dimension for different values of $Q$'s due to the limitation of computing time. But we have noticed that the greater the cutoff dimension is, the closer the frequencies are to the values in the RN case. Therefore, the results are definitely reliable.
    \begin{table}[h!]
    	\begin{center}
    		\begin{tabular}{|c|cc|cc|}
    			\hline
    			
    			$Q/M$ & \multicolumn{2}{c|}{\textbf{Charged KR BH}} & \multicolumn{2}{c|}{\textbf{RN BH}}\\
    			\hline
    			& $\omega_R M$ & $\omega_I M$ & $\omega_R M$ & $\omega_I M$\\
    			\hline
    			0 & 0.37361 & -0.088904 & 0.37367 &  -0.088962\\
    			\hline
    			0.2 & 0.37587 & -0.086122 & 0.37474 & -0.089075\\
    			\hline
    			0.4 & 0.37872 & -0.088766 & 0.37844 & -0.089398\\
    			\hline
    			0.6 & 0.38625 & -0.089803 & 0.38622 & -0.089814\\
    			\hline
    			0.8 & 0.40125 & -0.089731 & 0.40122 & -0.089643\\
    			
    			\hline
    		\end{tabular}
    	    \caption{The fundamental QNFs of gravitational perturbation with $L=2$ in the RN limit.}
    	    \label{CF1}
    	\end{center}
    \end{table}
    \begin{table}[h!]
    	\begin{center}
    		\begin{tabular}{|c|cc|cc|}
    			\hline
    			
    			$Q/M$ & \multicolumn{2}{c|}{\textbf{Charged KR BH}} & \multicolumn{2}{c|}{\textbf{RN BH}}\\
    			\hline
    			& $\omega_R M$ & $\omega_I M$ & $\omega_R M$ & $\omega_I M$\\
    			\hline
    			0 & 0.45758 & -0.095026 & 0.45759 & -0.095004\\
    			\hline
    			0.2 & 0.46266 & -0.093768 & 0.46297 & -0.095373\\
    			\hline
    			0.4 & 0.47986 & -0.096127 & 0.47993 & -0.096442\\
    			\hline
    			0.6 & 0.51201 & -0.098007 & 0.51201 & -0.098017\\
    			\hline
    			0.8 & 0.57014 & -0.099059 & 0.57013 & -0.099069\\
    			
    			\hline
    		\end{tabular}
    	    \caption{The fundamental QNFs of electromagnetic perturbation with $L=2$ in the RN limit.}
    	    \label{CF2}
    	\end{center}
    \end{table}
    
    As already discussed, suffering from numerical accuracy of continued fraction method\footnote{Restricting the calculation time within one day, the results only converge to two decimal places.}, we turn to the direct integration method on the study of the effect of LV on QNMs.
    %lay emphasis totally on the classification, stress this is only some calculation results, at the same time point out the problem.
    First, we test our discriminating method by eight different modes in the RN case with $Q=0.1M$, $M=1$ in Table~\ref{test}, where the third columns represented by $|C^{(a,b)}_1|/|C^{(a,b)}_2|$ shows the feature of the eigenstates as mentioned in Sec.~\ref{recog}. The ratio of the norm of two components of the eigenvalues clearly separate into two categories. The differences within each column are caused by numerical error of the state, since the perturbation equations in the RN case can be decoupled. It is worth mentioning that the ratio depends on the parameters. The reasons of the influence on the ratio caused by $Q$'s and $l$'s are different. $Q$'s represent different decouplable RN cases which means different transformation matrix, while $l$ represents only small deviation from the RN cases.
    
    \begin{table}[h!]
    	\begin{center}
    		\begin{tabular}{|ccc|ccc|}
    			\hline
    			
    			\multicolumn{3}{|c|}{\textbf{Gravitational}} & \multicolumn{3}{|c|}{\textbf{Electromagnetic}}\\
    			\hline
    			$\omega_R M$ & $\omega_I M$ &$|C^{(a,b)}_1|/|C^{(a,b)}_2|$ & $\omega_R M$ & $\omega_I M$ &$|C^{(a,b)}_1|/|C^{(a,b)}_2|$\\
    			\hline
    			0.373294  & -0.0886603 &5.78064 & 0.458462 & -0.0949216 &0.0310748\\
    			\hline
    			0.293899 & -0.136185 &4.88028 & 0.10007 &-0.112414 &0.0099893\\
    			\hline
    			0.389113 & -0.150272 &6.2845 & 0.295479  & -0.143227 &0.0217944\\
    			\hline
    			&  & & 0.631142 & -0.211206 &0.0441741\\
    			\hline
    			&  & & 0.708687 & -0.235345 &0.0495635\\
    			
    			\hline
    		\end{tabular}
    		\caption{Testing the feature of eigenvectors in the RN case with $Q=0.1M$ and $M=1$.}
    		\label{test}
    	\end{center}
    \end{table}
    
    Table~\ref{DI} shows the fundamental QNFs of a charged KR black hole with different values of the LV parameter $l$. Our calculations are carried out with $Q=0.1M$ and $M=1$. Here we give the trail solution\footnote{To find a root using numerical method would always need a trail value as an input.} in an iterative way. The results of frequencies are not very sound, the reason of which is the selection of parameters, as we would see in a while when doing error analysis. However, these results help in confirming our classification of modes. The separation of our discriminant stands as well, and the deviation of it from the RN case grows with the violation parameter $l$ increasing, although the deviation is relatively small. All the results support our analysis of the method of recognizing eigenvalues by investigating the eigenstates. Every following data will be tested by their discriminants. 
    
    \begin{table}[h!]
    	\begin{center}
    		\begin{tabular}{|c|ccc|ccc|}
    			\hline
    			
    			$l$ & \multicolumn{3}{c|}{\textbf{Gravitational}} & \multicolumn{3}{c|}{\textbf{Electromagnetic}}\\
    			\hline
    			& $\omega_R M$ & $\omega_I M$ &$|C^{(a,b)}_1|/|C^{(a,b)}_2|$ & $\omega_R M$ & $\omega_I M$ &$|C^{(a,b)}_1|/|C^{(a,b)}_2|$\\
    			\hline
    			0.01 & 0.3527 & -0.0818975 &5.12647 & 0.493944 & -0.0856006 &0.035185\\
    			\hline
    			0.02 & 0.35521 & -0.07568 &4.77627 & 0.502749 & -0.0792386 &0.0378984\\
    			\hline
    			0.03 & 0.35896 & -0.0723078 &4.43305 & 0.51033 & -0.0755524 &0.0408811\\
    			\hline
    			0.04 & 0.363159 & -0.0700878 &4.08147 & 0.517542 & -0.072982 &0.0442238\\
    			\hline
    			0.05 & 0.367616 & -0.0684941 &3.7184 & 0.524628 & -0.0710311 &0.048008\\
    			\hline
    			0.06 & 0.372263 & -0.0672944 &3.34354 & 0.531693 & -0.0694769 &0.0523307\\
    			\hline
    			0.07 & 0.377067 & -0.0663662 &2.95818 & 0.538792 & -0.0681999 &0.0573151\\
    			\hline
    			0.08 & 0.382014 & -0.065638 &2.56524 & 0.54596 & -0.0671289 &0.0631224\\
    			\hline
    			0.09 & 0.387095 & -0.0650669 &2.17 & 0.55322 & -0.0662177 &0.0699688\\
    			\hline
    			0.1 & 0.392308 & -0.0646298 &1.78149 & 0.560591 & -0.0654352 &0.0781509\\
    			
    			\hline
    		\end{tabular}
    		\caption{Testing the feature of eigenvectors in KR cases.}
    		\label{DI}
    	\end{center}
    \end{table}
    
    %下面被注释掉
    \iffalse
    \begin{table}[h!]
    	\begin{center}
    		\begin{tabular}{|c|ccc|ccc|}
    			\hline
    			
    			$l$ & \multicolumn{3}{c|}{\textbf{Gravitational}} & \multicolumn{3}{c|}{\textbf{Electromagnetic}}\\
    			\hline
    			& $\omega_R$ & $\omega_I$ &$|C^{(a,b)}_1|/|C^{(a,b)}_2|$ & $\omega_R$ & $\omega_I$ &$|C^{(a,b)}_1|/|C^{(a,b)}_2|$\\
    			\hline
    			0.01 & 0.403309 & -0.0832051 &5.92935 & 0.493944 & -0.0856006 &0.035185\\
    			\hline
    			0.02 & 0.412477 & -0.0769581 &5.72563 & 0.502749 & -0.0792386 &0.0378984\\
    			\hline
    			0.03 & 0.420189 & -0.073211 &5.48366 & 0.51033 & -0.0755524 &0.0408811\\
    			\hline
    			0.04 & 0.427472 & -0.0705365 &5.21674 & 0.517542 & -0.072982 &0.0442238\\
    			\hline
    			0.05 & 0.434609 & -0.0684726 &4.92864 & 0.524628 & -0.0710311 &0.048008\\
    			\hline
    			0.06 & 0.441721 & -0.0668102 &4.6209 & 0.531693  & -0.0694769 &0.0523307\\
    			\hline
    			0.07 & 0.448869 & -0.0654375 &4.29422 & 0.538792 & -0.0681999 &0.0573151\\
    			\hline
    			0.08 & 0.456092 & -0.0642879 &3.94893 & 0.54596 & -0.0671289 &0.0631224\\
    			\hline
    			0.09 & 0.463411 & -0.0633188 &3.58532 & 0.55322 & -0.0662177 &0.0699688\\
    			\hline
    			0.1 & 0.470842 & -0.0625014 &3.20375 & 0.560591 & -0.0654352 &0.0781509\\
    			
    			\hline
    		\end{tabular}
    		\caption{Testing the feature of eigenvectors in KR cases}
    		\label{DI}
    	\end{center}
    \end{table}
    \fi
    
    %第一段不改，第二段提到问题后说明这里试探解的取法和别处不一样
    %spiral line的说明中：此后均固定初值
    %加到绿线的说明中：此处l也在变化，但总体取值足够小，因此仍固定初值
    
    We mentioned above the data in Table~\ref{DI} are somewhat unreliable, this is implied at first glance by its discontinuity with the fundamental QNFs in the RN case. We also encountered further specific issues. All these issues stem from three influence factors. First, the values of violation parameter $l$ are too large. Physically, the violation of Lorentz symmetry cannot be apparent because of the broad support of general relativity from various tests, which means $l$ must be extremely small. In a bumblebee-induced model, $l$ has already been limited at the level of $10^{-13}$~\cite{restriction}. The problem of discontinuity can be avoided by choosing sufficiently small values of $l$. The second factor lies in the expansion order at boundary. The final factor is the position of boundary. When taking different values of $r_\mathrm{inf}$, we obtain different results. For example, we take the trail frequency $\omega_\mathrm{trail}M=0.388399-0.0731774i$ given by the continued fraction method which belongs to the gravitational category in RN case, the results are
    \begin{equation}
    	\begin{aligned}
    		\omega M&=0.403309-0.0832051i,\qquad|C^{(a,b)}_1|/|C^{(a,b)}_2|=5.92935,\qquad r_\mathrm{inf}=30,\\
    		\omega M&=0.383221-0.0366165i,\qquad|C^{(a,b)}_1|/|C^{(a,b)}_2|=0.0275663
    		,\quad r_\mathrm{inf}=100.
    	\end{aligned}
    \end{equation}
    We can see that the first result still belongs to the gravitational category, while the second one belongs to the electromagnetic category. Because the wave function will diverge at infinity, so $r_\mathrm{inf}$ should be chosen a proper value rather than a larger one.
    
\begin{figure}[htbp]
    	\centering
    	\begin{subfigure}[t]{0.48\textwidth}
    		\centering
    		\includegraphics[scale=0.7]{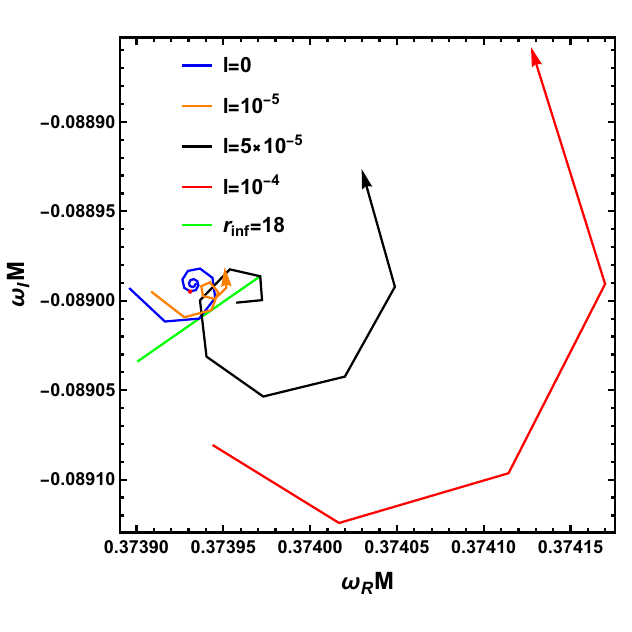}
    		\caption{Influence on the fundamental gravitational QNF by the violation parameter $l$ and boundary cutoff $r_\mathrm{inf}$ under 7 order expansion at both boundary ends.}
    		\label{circle1}
    	\end{subfigure}
    	\hfill
    	\begin{subfigure}[t]{0.48\textwidth}
    		\centering
    		\includegraphics[scale=0.7]{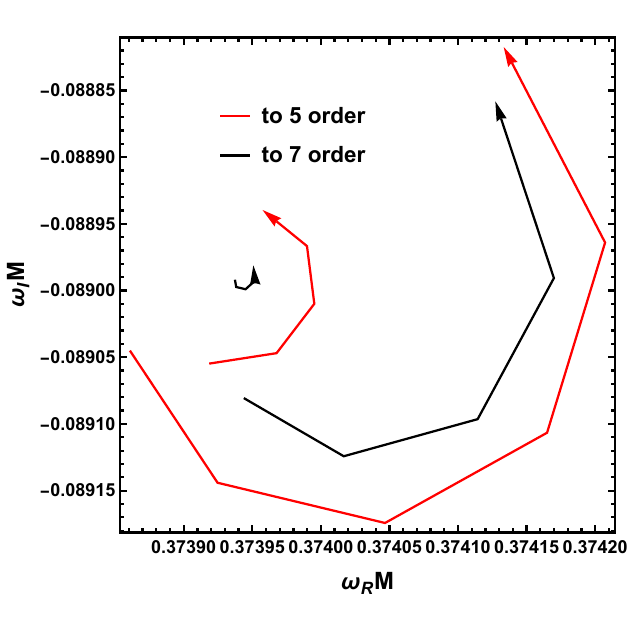}
    		\caption{Comparison between different order of expansion at both boundary ends.}
    		\label{circle2}
    	\end{subfigure}
    	\caption{Error analysis.}
    	\label{circle}
    \end{figure}

    From now on, we take the same trail solution $\omega_\mathrm{trail}M=0.373294-0.08866\mathrm{i}$, which is the fundamental gravitational QNF of RN black hole with $Q/M=0.1$. Figure~\ref{circle} shows the results of our error analysis. This is a scatter plot, and we connect the data points with lines for convenience. Different spiral lines represent different violation parameters, while each point on a line corresponds to a different $r_\mathrm{inf}$ value, which increases in the direction of arrows. The blue line represents the RN case and $r_\mathrm{inf}$ increases anticlockwisely which is similar to the other spiral lines. The cutoff ranges from 16 to 35 on the blue line, on which the red point represents $r_\mathrm{inf}=25$. On the other lines $r_\mathrm{inf}$ ends at 25 where the arrows are placed. The green line is obtained by changing $l$ fixing $r_\mathrm{inf}=18$, the effect of which is to be explained later. We do not take the trail solution in a iterative way as in Table.~\ref{DI} because the values of $l$ we take are adequately small in general.
     
\begin{figure}[htbp]
    	\centering
    	\begin{subfigure}[t]{0.48\textwidth}
    		\centering
    		\includegraphics[scale=0.7]{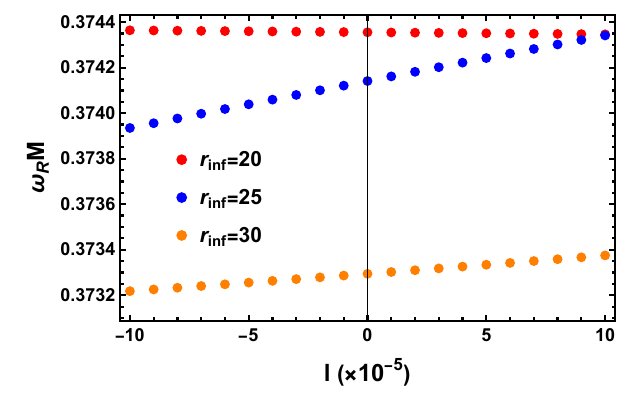}
    		\caption{Real part under 3 order expansion at horizon and 5 order expansion at infinity.}
    		\label{trend53re}
    	\end{subfigure}
    	\hfill
    	\begin{subfigure}[t]{0.48\textwidth}
    		\centering
    		\includegraphics[scale=0.7]{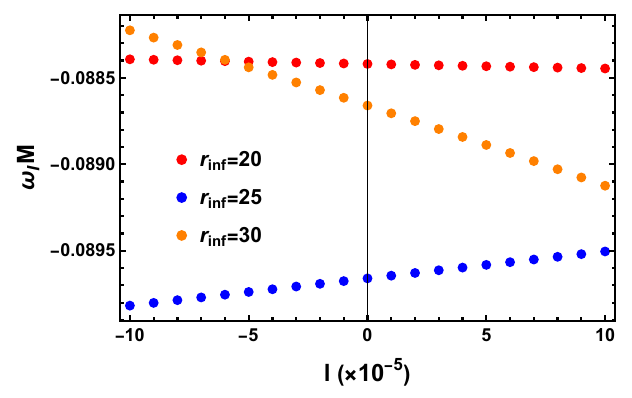}
    		\caption{Imaginary part under 3 order expansion at horizon and 5 order expansion at infinity.}
    		\label{trend53im}
    	\end{subfigure}
    	\vspace{0.8ex}
    	\begin{subfigure}[t]{0.48\textwidth}
    		\centering
    		\includegraphics[scale=0.7]{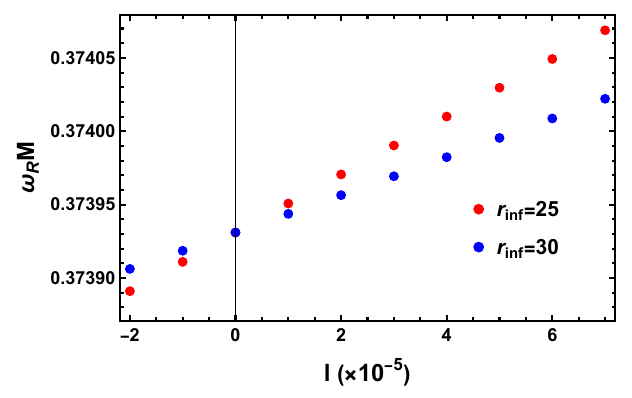}
    		\caption{Real part under 7 order expansion at both boundary ends.}
    		\label{trend7re}
    	\end{subfigure}
    	\hfill
    	\begin{subfigure}[t]{0.48\textwidth}
    		\centering
    		\includegraphics[scale=0.7]{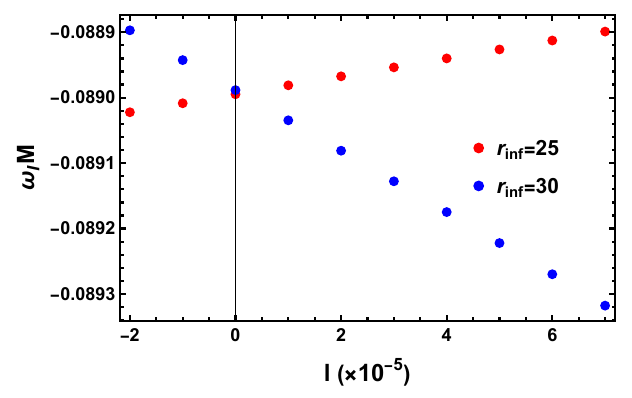}
    		\caption{Imaginary part under 7 order expansion at both boundary ends.}
    		\label{trend7im}
    	\end{subfigure}
    	\caption{The trend of gravitational fundamental QNF with respect to violation parameter $l$ under different boundary cutoff $r_\mathrm{inf}$.}
    	\label{trend}
    \end{figure}

    As can be easily read from the figure, all the results follow a similar spiral behavior. With relatively appropriate choice of $l$, there exist an ``innest circle'' which implies that in some region of $r_\mathrm{inf}$ we can get a steady solution. Away from the ``innest circle''s, the scale of spiral lines increases with $l$, while the ``phase'' of which does not change. The result is that even when fixing an inappropriate value of cutoff $r_\mathrm{inf}$, one would always get a linear behavior between $\omega$ and $l$ which seems to be a good result for some careless people. But actually it is not, and when choosing even neighboring $r_\mathrm{inf}$'s, one would find very different trends. This is more intuitive in Fig.~\ref{trend}, the trend can even change from increasing to decreasing with $l$. Surely, this analysis only stand when the parameters chosen are approximately appropriate, otherwise, one would find solutions in the wrong category. Note that all points here being recognized as belonging to the expected category can be viewed as another criterion that our results are reliable. After all those analyses, it is obvious that we should focus on the ``innest circle'' of each fixed $l$. Note also that the accuracy of a single point is actually adequately well, it is just not enough when studying the effect of $l$.
    
\begin{figure}
    	\centering
    	\includegraphics[scale=0.7]{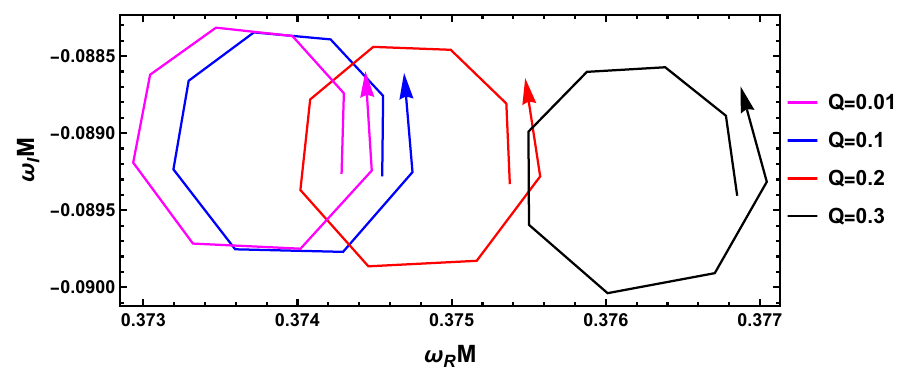}
    	\caption{Error analysis in RN cases under 3 order expansion at horizon and 5 order expansion at infinity.}
    	\label{RN}
    \end{figure}

    We have to mention that the same problem does not exist in RN case. Figure~\ref{RN} shows that the spiral lines obtained under different $Q$'s are of the same scale and ``phase''. As a result, for any fixed cutoff of appropriate magnitude, results do reflect both real and imaginary part behaviors.
    
\begin{figure}[htbp]
    	\centering
    	\begin{subfigure}[t]{0.48\textwidth}
    		\centering
    		\includegraphics[scale=0.7]{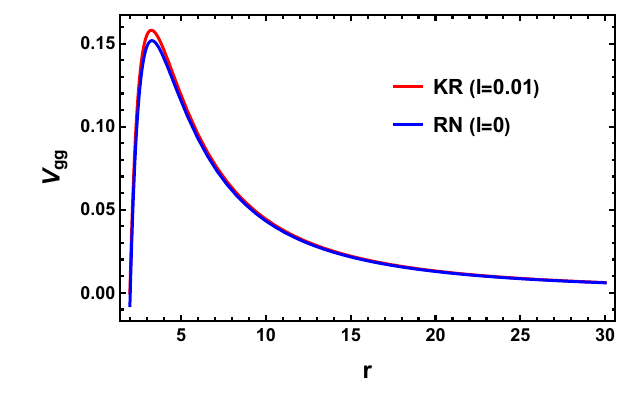}
    		\label{1}
    	\end{subfigure}
    	\hfill
    	\begin{subfigure}[t]{0.48\textwidth}
    		\centering
    		\includegraphics[scale=0.7]{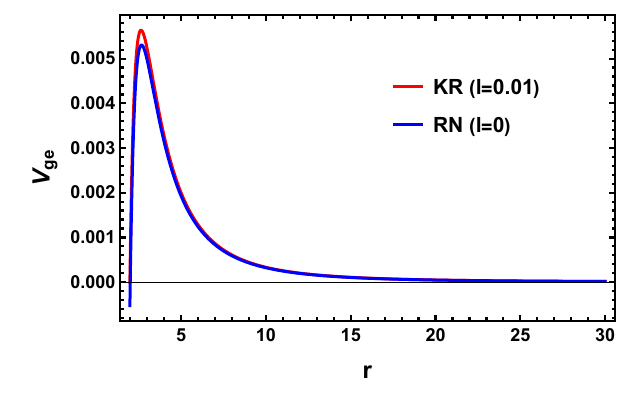}
    		\label{2}
    	\end{subfigure}
    	\vspace{0.8ex}
    	\begin{subfigure}[t]{0.48\textwidth}
    		\centering
    		\includegraphics[scale=0.7]{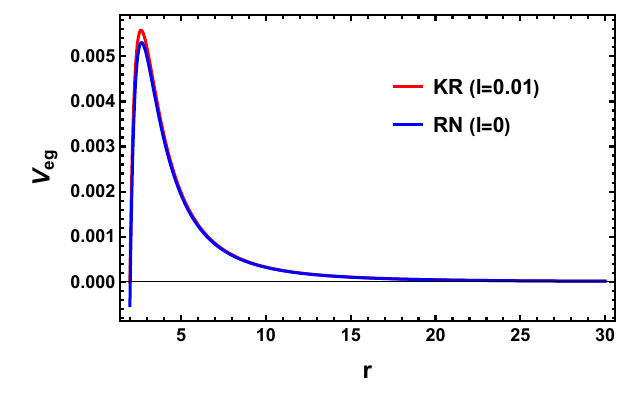}
    		\label{3}
    	\end{subfigure}
    	\hfill
    	\begin{subfigure}[t]{0.48\textwidth}
    		\centering
    		\includegraphics[scale=0.7]{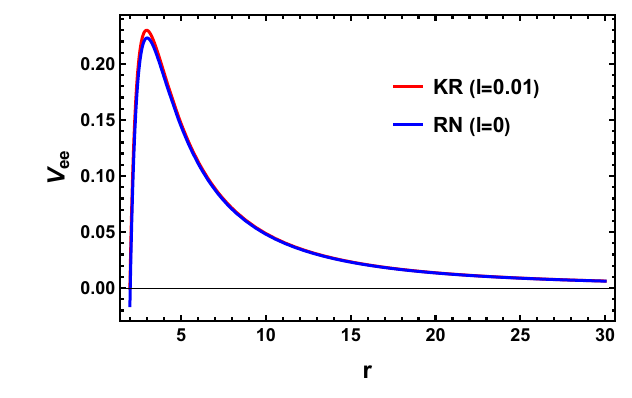}
    		\label{4}
    	\end{subfigure}
    	\caption{Effective potential components compared with RN case.}
    	\label{potential1}
    \end{figure}

    Another thing to mention, from the analyses above, the impact of cutoff $r_\mathrm{inf}$ on the results is very significant. We have to emphasize that the influence of cutoff is caused by wavefunction-divergence-induced numerical overflow instead of the deviation of effective potentials shown in Figs.~\ref{potential1} and \ref{potential2}. From the graphs it is easy to see that the influence on effective potential by violation parameter $l$ is very small relatively.
    
\begin{figure}[htbp]
    	\centering
    	\includegraphics[scale=0.7]{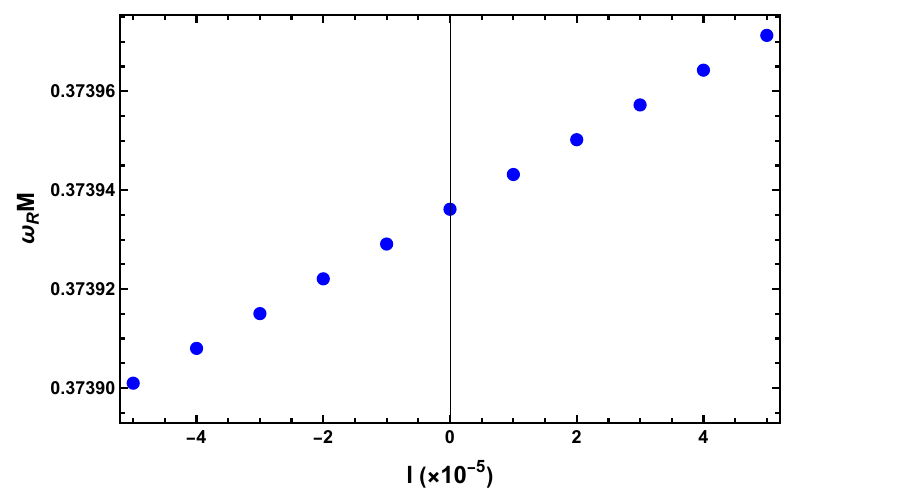}
    	\caption{Effect of violation parameter $l$ on the real part of gravitational fundamental QNF with boundary cutoff $r_\mathrm{inf}=18$ under 7 order expansion at both boundary ends.}
    	\label{result}
    \end{figure}

    We then study the movement of the ``innest circle''. Figure~\ref{circle2} shows the improvement brought by increasing the order of expansion at boundary, but the calculating time increases rapidly. Since we aim to investigate how the ``innest circle'' behaves with respect to violation parameter, it is beneficial to choose as possible higher order of expansion. Alternatively, at the very least, the analysis should be conducted over a wider range of $l$. This is because using a lower order of expansion might cause multiple ``innest circle''s to nest within the same region of $l$. We finally choose 7 order of expansion, controlling the calculating time of one point to about one day on a personal computer. Again, in Fig.~\ref{circle1}, the key difference among the spiral lines is the region near ``innest circle''s. The corresponding $r_\mathrm{inf}$ of the ``innest circle'' for different spiral lines decreases with $l$ increasing. So theoretically there is no appropriate choice of a fixed $r_\mathrm{inf}$. For each $l$, the ``innest circle'' has to be found, but this would require too much time. Observing the left three spiral lines in Fig.~\ref{circle1}, the ``innest circle''s almost move horizontally to the right. This means that the real parts of fundamental mode increase obviously with $l$, while the imaginary parts stay unchanged, or at least insufficient to be studied under current level of accuracy. Now we focus on the green line, $r_\mathrm{inf}=18$ is on the ``innest circle''s of some spiral lines when $l$ is relatively larger such as on the black one, but for other spiral lines, for example the orange and blue ones, the real parts of the points on the green line are almost the same as the real parts of the corresponding ``innest circle''s. So we calculate the trend of real parts of gravitational fundamental frequencies with the green line as shown in Fig.~\ref{result}. This is a little opportunistic but reasonable way we apply. Of course we view the imaginary parts of which as not changing, and can be set directly as the frequency in the RN case. It shows that the real parts increase linearly with $l$ at least within a proper region of $l$.
    
    As mentioned above, the LV parameter $l$ in a bumblebee-induced model has already been limited at the level of $10^{-13}$~\cite{restriction}. Since the violations in both the Bumblebee- and KR-induced models are caused by spontaneous LV, the restriction can definitely be referred to here in our case. Because continue lowering $l$ may increase the calculating time significantly, our choice of $l$ is not sufficiently small. But we can reasonably infer from the linear behavior we obtained of the real part and the invariance of the imaginary part the fundamental gravitational frequencies when $l$ is set less then $10^{-13}$.
    
    %The last thing to mention, we only resolves the problem of recognizing whether a mode is gravitational or electromagnetic, this surely helps a lot on our analyses, but it can still not tell whether a mode is fundamental, or more generally, whether a mode correspond to the same mode as our trail solution, or even is it a new mode. In our analyses of fundamental mode, the results can still be viewed effective by its continuity with RN fundamental mode, but analyze the whole spectrum of this system would need more efforts.
    
    \iffalse
    \begin{figure}[htbp]
    	\centering
    	\begin{subfigure}[t]{0.24\textwidth}
    		\centering
    		\includegraphics[scale=0.45]{Vgg}
    		\label{1}
    	\end{subfigure}
    	\hfill
    	\begin{subfigure}[t]{0.24\textwidth}
    		\centering
    		\includegraphics[scale=0.45]{Vga}
    		\label{2}
    	\end{subfigure}
    	\hfill
    	\begin{subfigure}[t]{0.24\textwidth}
    		\centering
    		\includegraphics[scale=0.45]{Vag}
    		\label{3}
    	\end{subfigure}
    	\hfill
    	\begin{subfigure}[t]{0.24\textwidth}
    		\centering
    		\includegraphics[scale=0.45]{Vaa}
    		\label{4}
    	\end{subfigure}
    	\vspace{0.8ex}
    	\begin{subfigure}[t]{0.32\textwidth}
    		\centering
    		\includegraphics[scale=0.5]{deltaV_l=0.01}
    		\label{5}
    	\end{subfigure}
    	\hfill
    	\begin{subfigure}[t]{0.32\textwidth}
    		\centering
    		\includegraphics[scale=0.5]{deltaV_l=0.0001}
    		\label{6}
    	\end{subfigure}
    	\hfill
    	\begin{subfigure}[t]{0.32\textwidth}
    		\centering
    		\includegraphics[scale=0.5]{deltaV_l=-0.01}
    		\label{7}
    	\end{subfigure}
    	\caption{The effective potential and its deviation from RN case.}
    	\label{potential}
    \end{figure}
    \fi
    
    %\iffalse
    %\iffalse
    \begin{figure}[htbp]
    	\begin{subfigure}[t]{0.48\textwidth}
    		\centering
    		\includegraphics[scale=0.7]{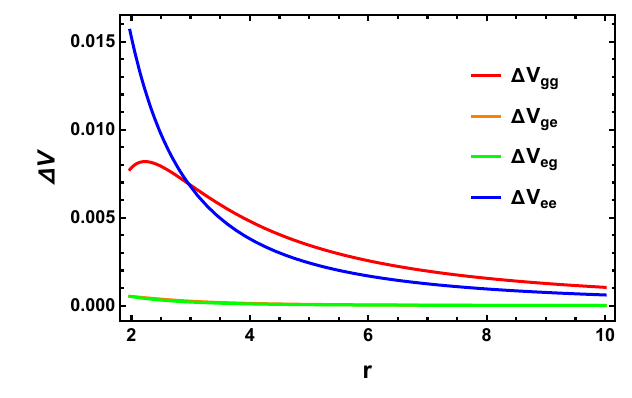}
    		\caption{violation parameter $l=0.01$.}
    		\label{5}
    	\end{subfigure}
    	\hfill
    	\begin{subfigure}[t]{0.48\textwidth}
    		\centering
    		\includegraphics[scale=0.7]{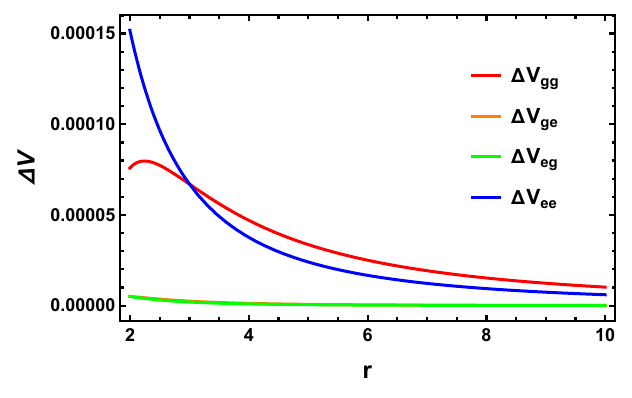}
    		\caption{violation parameter $l=0.0001$.}
    		\label{6}
    	\end{subfigure}
    	\vspace{0.8ex}
    	\begin{subfigure}[t]{\textwidth}
    		\centering
    		\includegraphics[scale=0.7]{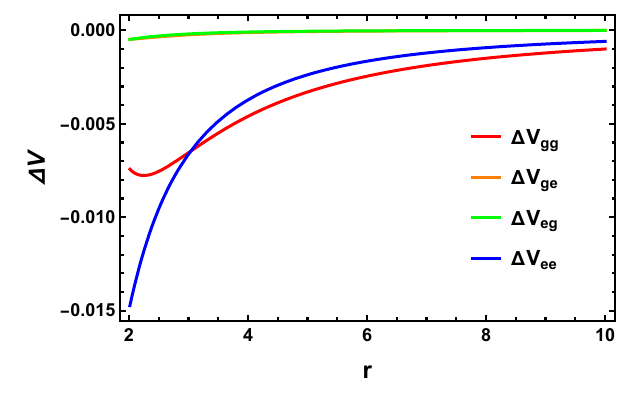}
    		\caption{violation parameter $l=-0.01$.}
    		\label{7}
    	\end{subfigure}
    	\caption{Deviation of effective potential components from RN case at different violation parameter $l$.}
    	\label{potential2}
    \end{figure}
    %\fi
    \iffalse
    \begin{figure}[htbp]
    	\begin{subfigure}[t]{0.32\textwidth}
    		\centering
    		\includegraphics[scale=0.7]{deltaV_l=0.01_1a}
    		\caption{violation parameter $l=0.01$.}
    		\label{5}
    	\end{subfigure}
    	\hfill
    	\begin{subfigure}[t]{0.32\textwidth}
    		\centering
    		\includegraphics[scale=0.7]{deltaV_l=0.0001_1a}
    		\caption{violation parameter $l=0.0001$.}
    		\label{6}
    	\end{subfigure}
    	\begin{subfigure}[t]{0.32\textwidth}
    		\centering
    		\includegraphics[scale=0.7]{deltaV_l=-0.01_1a}
    		\caption{violation parameter $l=-0.01$.}
    		\label{7}
    	\end{subfigure}
    	\caption{Deviation of effective potential components from RN case at different violation parameter $l$.}
    	\label{potential2}
    \end{figure}
    \fi
    
\section{Conclusion}\label{Conclusion}
    LV is a possible Planck-scale signal of modified gravity. A natural way of generating it in string theory is spontaneous LV. The KR field, dual to the axion field, can induce this process.
    
    %work better as an abstract?
    In this paper, the perturbation problem of an electrically charged spherically symmetric KR black hole was studied. A method was developed to recognize the eigenvalues, especially for near decouplable equation sets. We performed a numerical error analysis of the matrix-valued direct integration method. The results help us choose the appropriate region of violation parameter and the boundary cutoff. Our calculation shows that the violation parameter mainly causes the deviation of real parts of the fundamental modes linearly with $l$, while the imaginary parts remain unchanged. We also presented the problems remained at the very end.
    
    It is worth mentioning that, since our research was based on a black hole spacetime purely caused by spontaneous LV, there is great probability that the behavior of QNFs presented here cannot be generalized to phenomenological researches including more general Lorentz-violating cases as mentioned in the introduction. So studies are on the ways to recognize different kinds of LV, or maybe a more general description of LV beyond SME are expected to be done.
    
    \section*{Acknowledgements}
    This work was supported by the National Natural Science Foundation of China (Grants No. 12475056, No. 12247101, No. 12205129, No. 12347111, and No. 12405055), the Fundamental Research Funds for the Central Universities (Grant No. lzujbky-2025-jdzx07), the Natural Science Foundation of Gansu Province (No. 22JR5RA389, No.25JRRA799), and the ‘111 Center’ under Grant No. B20063, Wen-Di Guo was supported by “Talent Scientific Fund of Lanzhou University”.

%\appendices
%\section{A}
	
	%\bibliographystyle{IEEEtran}
	%\bibliography{ref}

\end{document}